\documentclass[showpacs,nobibnotes,nofootinbib,preprint]{revtex4}
\usepackage{graphicx}
\usepackage{amsmath}
\usepackage{amsfonts}
\begin{document}
\newcommand{\beq}{\begin{equation}}
\newcommand{\eeq}{\end{equation}}
\newcommand{\barr}{\begin{eqnarray}}
\newcommand{\earr}{\end{eqnarray}}
\def\figwidth{7.5cm}
\newcommand{\ascomm}[1]{{\bf AS Comm: -#1- End of Comment}}
\newcommand{\asdel}[1]{[[ AS : #1 ]]}
\newcommand{\asdrop}[1]{[[ AS : #1 ]]}
\newcommand{\asadd}[1]{{\bf #1}}
\def\cH{{\cal H}}
\def\cP{{\cal P}}
\def\cD{{\cal D}}
\def\cG{{\cal G}}
\def\cV{{\cal V}}
\def\cF{{\cal F}}
\def\cU{{\cal U}}
\def\cS{{\cal S}}
\def\cM{{\cal M}}
\def\cO{{\cal O}}
\def\cE{{\cal E}}
\def\bfA{{\bf A}}
\def\bfG{{\bf G}}
\def\bfn{{\bf n}}
\def\bfr{{\bf r}}
\def\bfV{{\bf V}}
\def\bft{{\bf t}}
\def\bfM{{\bf M}}
\def\bfP{{\bf P}}
\def\Tr{{\rm Tr\ }}
\def\bra#1{\langle #1 |}
\def\ket#1{| #1 \rangle}
\newcommand{\p}{\partial}
\def\coltwovector#1#2{\left({#1\atop#2}\right)}
\def\upp{\coltwovector10}
\def\downn{\coltwovector01}
\def\Ord#1{{\cal O}\left( #1\right)}
\renewcommand{\Re}{{\rm Re}}
\renewcommand{\Im}{{\rm Im}}
\def\ask{\marginpar{?? ask:  \hfill}}
\def\fin{\marginpar{fill in ... \hfill}}
\def\note{\marginpar{note \hfill}}
\def\check{\marginpar{check \hfill}}
\def\discuss{\marginpar{discuss \hfill}}
\title{Casimir Dynamics: \\ Interactions of Surfaces with Codimension $> 1$ Due to Quantum Fluctuations}
\author{A.~Scardicchio}\email{scardicc@mit.edu}
\affiliation{Center for Theoretical Physics, \\ Laboratory for
   Nuclear Science and Department of Physics \\ Massachusetts
Institute of Technology \\ Cambridge, MA 02139, USA}
\begin{abstract}
    \noindent  We study the Casimir force between defects (branes) of co-dimension larger than 1
    due to quantum fluctuations of a scalar field $\phi$ living in the bulk.
    We show that the Casimir force is attractive and that it diverges
    as the distance between the branes approaches a critical value $L_c$. Below this critical
    distance $L_c$ the vacuum state $\phi=0$ of the theory is unstable, due to the birth
    of a tachyon, and the field condenses.
\end{abstract}
 \vspace*{-\bigskipamount}
\preprint{MIT-CTP-3615} \maketitle \setcounter{equation}{0}

\section{Introduction}

Point-like interactions have provided a remarkably useful
idealization for many situations in physics. In the context of
scattering theory the concept of a point-like scatterer was
introduced in 1934 by Bethe and Peierls \cite{Bethe}. Fermi
\cite{Fermi} used and refined their results to describe the motion
of neutrons in hydrogenated substances (such as paraffin) by
introducing what is now known as the `Fermi pseudopotential'. The
idea is that when the scattering potential is concentrated on a
very small scale $r_0$ (in the case studied by Fermi the range was
that of nuclear interactions compared to the distances between the
atoms) but nonetheless its influence on the motion cannot be
neglected, one can characterize the scattering in a simple and
efficient way by means of a few quantities like the scattering
length, finite in the $r_0\to 0$ limit. The problem of `how to
separate the scales' in the Schr\"odinger equation triggered by
those 1930's papers was addressed and elegantly solved over the
years at different levels of formalism
\cite{Zeldovich,Faddeev,Albeverio,Jackiw,Solodukhin1}. The key to
the solution relies in a proper definition of a `delta-function
interaction' in dimensions greater than 1.

Of course this paper will not be dealing with quantum mechanical
scattering within matter, which is from many points of view a
solved problem. The problem of `separation of scales' however
arises urgently in modern quantum field theory if stated as:
``What is the quantum field theory response on length-scales $L$
to a disturbance concentrated on a length scale $r_0\ll L$?" How
does quantum field theory respond to topological defects and
singularities, in particular of the metric? Once formalized in
proper mathematical terms the two problems look much closer than
one would think.

The Casimir effect falls in this class of problems. The
penetration length ($r_0$) of the electromagnetic field inside
conductors is much smaller than the distance ($L$) between the
conductors, which sets the scale of the experimentally measured
force. We are interested in studying the dynamics of the
conductors on the larger scale $L$ by integrating out the
electromagnetic field. This motivates the nomenclature `Casimir
effect' for a much wider set of problems than Casimir's original
one.

Another example: in any candidate theory of the quantum geometry
of space-time the problem of dealing with point-like singularities
will inevitably arise. Remember for example that the Ricci scalar
for a point-like particle (like Schwartzschild's solution) is a
delta-function centered on the position of the particle. Quadratic
fluctuations of a non minimally-coupled scalar (or of the metric)
have hence a delta-function term in their Lagrangian. The effect
of such a term must be considered together with the other known
effects of the black hole metric. It is then of paramount
importance to analyze the problem of how one or more concentrated
singularities influence the spectrum and low-energy behavior of
the fluctuations of the field.

Analogous problems arise in condensed matter, quantum field theory
and string theory since localized disturbances appear in all these
theories, essentially only the names given to them being different
(defects, domain walls, concentrated Aharonov-Bohm fluxes and
branes to name some). With this in mind we will set-up the problem
in very general terms and even though not all the details map
one-to-one on specific examples the main results will apply to a
wide class of examples.

The main results of this paper are two. First, I will show that
quantum fluctuations of a scalar field $\phi$ generate attractive
forces between localized defects in the very same way Casimir
forces act between metallic bodies. I calculate this force for an
arbitrary number of defects with
co-dimension\footnote{Co-dimension is the number of dimensions
transverse to a manifold. For example a point in 2 dimensions, a
line in 3 dimensions and a surface in 4 dimensions all have
co-dimension 2.} 1, 2 and 3 (see Table I). Note that previously
the Casimir effect has been analyzed only for co-dimension 1. The
main result of this paper is Eq.~(\ref{eq:Eintwickrot}), which
gives the interaction energy as a function of the scattering
lengths of the defects and their relative separations. For
co-dimension $d\geq 4$ the force disappears, as required by the
properties of the self-adjoint extensions of the Laplace operator
on the punctured $\mathbb{R}^{d}$ (see \cite{Simon}, Chap.~X).

Second, in the presence of two or more of these defects (of
co-dimension $>1$) the vacuum $\phi=0$ is hopelessly unstable and
a localized tachyon mode is formed when the defects approach
closer than a critical distance. At this critical distance the
attractive force diverges. I calculate the wave function of the
tachyon and show that it leads to condensation of the bulk field
$\phi$ to a vacuum expectation value $\overline{\phi}(x)\neq 0$
---but only in a limited region of space.
The consequences of these observations for some models will be
discussed in Section \ref{sec:omiss}.

\begin{table}[t] 
   \centering
   \includegraphics[scale=1]{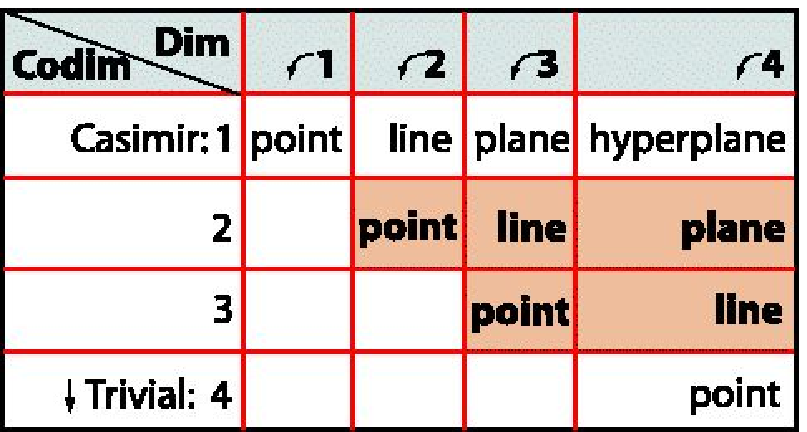}
   \caption{Flat manifolds divided according to dimension and
co-dimension. The first line is the well-known Casimir problem,
from the fourth line down the perturbation is `invisible' to
fluctuations. In this paper we will be dealing with manifolds in
lines 2 and 3.}
   \label{fig:example}
\end{table}

\section{The interaction energy}

In this Section we will calculate the effective action
\cite{Weinberg} of a scalar field coupled quadratically to a
static configuration of defects. The effective action $S_{{\rm
eff}}$ and Casimir energy $\cE$ are proportional to each other
\beq
S_{{\rm eff}}=-T \cE,
\eeq
where $T$ is the interaction time. In the following we will be
interested in the Casimir energy of the problem. We will see that
the part of the Casimir energy responsible for the interaction
between the defects is a cutoff-independent quantity, meaning that
the separation of scales can be performed effectively in this
quantum field theory.

We will consider the following action for the scalar field in
$d+1$ dimensions ($\hbar=c=1$):
\beq
\label{eq:Sphi}
S_\phi=\int d^dx dt\ \frac{1}{2}(\partial
\phi)^2-\frac{1}{2}\left(m^2+\sum_i^N\mu_i\delta(x-a_i)\right)\phi^2.
\eeq
Here $\delta$ is the $d-$dimensional Dirac's delta-function,
mimicking the concentrated disturbance on the field, $\mu_i$ are
constants, meaning that they do not depend on the field $\phi$,
but in general they can depend functionally on other fields living
on the defect.\footnote{The $\mu_i$'s are also called `brane
tensions'.} The methods of \cite{Albeverio} will be used in order
to define these $\delta$'s correctly. In this section, in order to
keep things simple we restrict our attention to points in 1, 2 and
3 dimensions. We will add an arbitrary number of flat directions
in Section \ref{sec:Extn}, hence fulfilling our promise of
studying co-dimensions 1, 2 and 3.

Actions like Eq.~(\ref{eq:Sphi}) arise in different contexts. For
example consider the case of a scalar bulk field $\phi$ coupled
with $N$ branes in curved or flat space (curvature can be easily
included in Eq.~(\ref{eq:Sphi})); or the case of a cosmic string
(again the curvature outside the string must be considered); or
the case of electrons coupled with Aharonov-Bohm fluxes (in this
case one has fermions rather than bosons but, after squaring the
Dirac equation, the analysis is analogous \cite{Jackiw}). All
these examples can be studied with the formalism introduced in
this paper, so in full generality we will study the quantum
fluctuations of the action Eq.~(\ref{eq:Sphi}).

Some features of the Eq.~(\ref{eq:Sphi}) with only one delta
function and constant $m$ have been studied before, for example in
connection with cosmic strings scenarios
\cite{Solodukhin1,Allen,Kay}. The action (\ref{eq:Sphi}) in one
dimension with a single delta and a space-dependent mass term
$m^2=m^2(x)$ has been studied in \cite{Buoyancy}.\footnote{The
scale of variation of $m^2$ should be of the order of the `long
scale' $L$. The purpose of this paper is to integrate out the
physics at momenta $\gtrsim 1/r_0$ which is symbolized by the
delta functions in Eq.~(\ref{eq:Sphi}).} In this paper we will
consider the situation where a generic number of defects are
present in $d\geq 1$ and $m^2$ is a constant. We will see that the
situation will be different from that depicted in
Ref.s~\cite{Solodukhin1,Allen,Buoyancy} and unexpected physics is
found. Moreover the generalization to $x$-dependent $m^2$ can be
easily achieved by means of the techniques of Ref.~\cite{Buoyancy}
and we will not comment on it in this paper. The inclusion of any
other term in the action (\ref{eq:Sphi}) describing the dynamics
of the surface itself would not affect our calculations. We
consider the positions of the defects $a_i$ fixed and obtain an
effective action. In the usual way this action can be used to
describe adiabatically moving $a_i(t)$ (\emph{i.e.}\ if the
velocities $|\dot{a}_{ij}|\ll c$).

The forces between the defects can be calculated by taking the
derivatives of the Casimir energy $\cE$ with respect to its
arguments $\{a_i\}$. It will turn out that in general the forces
are not additive, \emph{i.e.}\ $\cE$ is not a superposition of
terms depending only on the relative distances $a_{ij}\equiv
|a_i-a_j|$.

We use the following integral representation of the zero-point
energy of the scalar field $\phi$,
\beq
\label{def:CE}
\cE=\frac{1}{2}\int_0^\Lambda dE\rho(E)\sqrt{E},
\eeq
where $\Lambda$ is a cutoff and $\rho$ is the spectral density of
the Hamiltonian operator $H$
\beq
\label{eq:H}
H=-\nabla^2+m^2+\sum_{i=1}^N\mu_i\delta(x-a_i).
\eeq
Define also the \emph{unperturbed} hamiltonian $H_0$ as
\beq
\label{eq:H0}
H_0=-\nabla^2+m^2.
\eeq
Equation (\ref{def:CE}) will look more familiar if the replacement
$E\to \omega^2$ is done, and $\hbar$'s are restored. One can
obtain the spectral density $\rho(E)$ as a functional of the
propagator $G(E)=1/(H-E)$ as
\beq
\rho(E)=\frac{1}{\pi}\lim_{\epsilon\to 0^+}\Im\,\Tr
G(E+i\epsilon).
\eeq
In the following we will often write $E+i0^+$ for $E+i\epsilon$
when $\epsilon\to 0^+$.

The propagator, $\cG(x',x;E)\equiv\bra{x'}G(E)\ket{x}$ satisfies
the Schr\"odinger equation
\beq
(-\nabla'^2+m^2+\sum_{i=1}^N\mu_i\delta(x'-a_i)-E)\cG(x',x;E)=\delta(x'-x),
\eeq
and $\cG_0\equiv \bra{x'}G_0(E)\ket{x}$ satisfies the analogous
equation without $\delta$'s on the left-hand side. For $m$
constant (which we will assume unless explicitly stated) and $\Im
E>0$ we have\footnote{In the following we will not use any special
notation for vectors and we will indicate with $|x|$ the norm of a
vector in 1,2 and 3 dimensions.}
\beq
\cG_0(x',x;E)=
\begin{cases}
\frac{i}{2\sqrt{E-m^2}}e^{i\sqrt{E-m^2}|x'-x|} &\text{if $d=1$}\\
\frac{i}{4}H_0^{(1)}(\sqrt{E-m^2}|x'-x|) &\text{if $d=2$}\\
\frac{e^{i\sqrt{E-m^2}|x'-x|}}{4\pi|x'-x|} &\text{if $d=3$},
\end{cases}
\eeq
where $H_0^{(1)}$ is Hankel's function of first kind of order 0.

For $d=1$ the problem is that of a scalar field on the line
$\mathbb{R}$ in the background of a stack of $\delta$-functions
centered on $x=\{a_i\}$ \cite{Jaffe}. If we assume $\mu_i>0$ they
will attract each other, like metallic plates do via the Casimir
effect. These forces are not confining and no new physics is
obtained with the generalization obtained by adding $n$ transverse
directions. This is the usual Casimir problem. We will see how the
situation changes dramatically when $d>1$.

To see how the solution for $\cG$ is obtained, consider first the
case with a single delta function with strength $\mu_1=\mu$,
placed at $x=a$. By solving the Lippman-Schwinger equation
\cite{Zeldovich,Buoyancy, Albeverio} one finds
\beq
\label{eq:1delta1d}
\cG(x',x;E)=\cG_0(x',x;E)+\frac{1}{\alpha-\cG_0(a,a;E)}\cG_0(x',a;E)\cG_0(a,x;E),
\eeq
where $\alpha=-1/\mu$. This solution is perfectly good in $d=1$
and was the basis of the analysis in \cite{Buoyancy} for
non-constant $m^2$. For $d>1$ it however suffers from a serious
problem since $\cG(a,a+r;E)\to\infty$ when the point splitting
regulator $r\equiv|r|\to 0$. One can reabsorb this divergence
\cite{Zeldovich} in a redefinition of $\alpha$ to obtain a finite
result
\beq
\label{eq:1delta1dren}
\cG(x',x;E)=\cG_0(x',x;E)+\frac{1}{\alpha_r-B^{(d)}(E)}\cG_0(x',a;E)\cG_0(a,x;E),
\eeq
where
\beq
\label{eq:Bd}
B^{(d)}=
\begin{cases}
\frac{i}{2\sqrt{E-m^2}} &\text{if $d=1$}\\
-\frac{1}{2\pi}\ln\left(\frac{\sqrt{E-m^2}}{iM}\right) &\text{if
$d=2$}\\
\frac{i\sqrt{E-m^2}}{4\pi} &\text{if $d=3$},
\end{cases}\\
\eeq
where for $d=2$ it has been necessary to introduce an arbitrary
mass scale $M$ which stays finite when the point splitting
regulator $r\to 0$. So $\alpha$ must be redefined such that when
$r\to 0$
\beq
\alpha_r=
\begin{cases}
\alpha &\text{if $d=1$}\\
\alpha + \frac{1}{2\pi}\ln M r &\text{if $d=2$}\\
\alpha - \frac{1}{4\pi r} &\text{if $d=3$},
\end{cases}
\eeq
so the `renormalized' $\alpha_r$ stays finite. It is clear from
this equation that one needs a \emph{positive} divergent $\alpha$
to reabsorb the negative divergences when $r\to 0$. Large positive
$\alpha$ mean \emph{negative} very small $\mu$ (since
$\alpha=-1/\mu$). A small negative $\mu$ corresponds to a weakly
attractive potential. Hence the point-like scatterer limit can be
thought of as the limit of a concentrated attractive potential,
zero outside a sphere of radius $r_0$, with at most one bound
state whose energy stays finite when $r_0\to 0$
\cite{Zeldovich,Faddeev}. For $d=2$ any attractive potential has
at least a bound state and so we always find a bound state also
for $r_0\to 0$ (for $d=2$ the dependence of $\alpha$ on $M$ is
reminiscent of a renormalization group flow
\cite{Jackiw,Allen,GW1}); for $d=3$ the bound state can be real or
`virtual' (\emph{i.e.}\ a pole of the propagator $G(E)$ located on
the second Riemann sheet) its energy being finite in the limit
$r_0\to 0$. The scattering length is (both for $d=2$ and $3$) a
function of $\alpha$ and is hence finite in the $r_0\to 0$ limit.

Another interpretation of this results comes from the theory of
self-adjoint extensions of symmetric operators
\cite{Faddeev,Albeverio}. Here the `renormalized' $\alpha_r$
corresponds to a choice of self-adjoint extension for the
Laplacian operator $-\Delta$ on the punctured $\mathbb{R}^d$
\cite{Faddeev}. In $\mathbb{R}^2$ the self-adjoint extensions are
not positive definite, meaning that they all have at least one
(but it turns out there is only one) negative eigenvalue. This
corresponds to the bound state described in the paragraph above.
In the punctured $\mathbb{R}^3$ the self-adjoint extensions of
$-\Delta$ can be either positive semi-definite (with a virtual
state on the second Riemann sheet) or not (due to the existence of
a single real and negative eigenvalue).

The propagator with $N$ deltas at positions $\{a_i\}$,
$i=1,...,N$, and $1\leq d \leq 3$ can be found \cite{Albeverio}:
\beq
\label{eq:GNdelta}
\cG(x',x;E)=\cG_0(x',x;E)+\sum_{i,j=1}^N(\Gamma^{-1})_{ij}\cG_0(x',a_i;E)\cG_0(a_j,x;E).
\eeq
The matrix $\Gamma$ is defined as (from now on we drop the
subscript $r$ on the $\alpha_r$'s)
\beq
\label{eq:gammad3}
\Gamma_{ij}=\left(\alpha_i-B^{(d)}(E)\right)\delta_{ij}-\widetilde{\cG}_0(a_i,a_j;E),
\eeq
where
\barr
\label{eq:tildeg0}
\widetilde{\cG}_0(a_i,a_j;E)&=&
\begin{cases}
0 &\text{if $i=j$}\\
\cG_0(a_i,a_j;E) &\text{if $i\neq j$}.
\end{cases}
\earr

It is now possible to explain why we limited our discussion to
$d\leq 3$. The reason is that looking at the Laplacian $\Delta$ on
the punctured $\mathbb{R}^4$ one realizes that this operator is
essentially self-adjoint \cite{Albeverio,Simon}, meaning that it
has a unique self-adjoint extension: the trivial one. The
4-dimensional delta function is `too small' a perturbation to be
seen by the Laplacian. What does go wrong in the renormalization
procedure? The propagator in $d=4$ is ($|x'-x|\equiv r$)
\barr
\cG_0(x',x;E)&=&\frac{i \sqrt{E-m^2}}{8\pi
r}H_1^{(1)}(\sqrt{E-m^2}\ r)\\
&\sim&\frac{i}{8\pi}\left(\frac{1}{r^2}+\sqrt{E-m^2}\ln
r+\Ord{1}\right)\quad \text{if $r\sim 0$}
\earr
so we cannot choose $\alpha$ in an $E$-independent way (because of
the $\sqrt{E-m^2}\ln r$ term) to remove completely the divergences
as we did before. The low-energy limit of a $4-$dimensional
concentrated potential is hence trivial and we will not discuss
this problem anymore.

Having solved for the propagator we can find the density of states
$\rho$ simply by taking the trace and the imaginary part. The
result\footnote{The only algebraic identity worth of notice is the
fact that
\beq
\sum_{ij}(\Gamma)^{-1}_{ij}\frac{\partial}{\partial
E}\cG_0(a_i,a_j;E)=-\frac{\partial}{\partial E}\Tr\ln\Gamma.
\eeq
 } is \cite{Buoyancy}
\beq
\label{eq:rho}
\rho(E)=\rho_0(E)-\frac{1}{\pi}\Im\frac{\partial}{\partial
E}\ln\det\Gamma(E+i 0^+,\{a\}),
\eeq
where $\rho_0=\pi^{-1}\Im\Tr G_0$ and the determinant of $\Gamma$
is simply the determinant over the matrix indices $ij$.

The term $\rho_0$ in (\ref{eq:rho}) is independent of the
presence, strengths $\alpha_i$ and relative positions of the delta
functions and we will neglect it in the following. The second term
on the right-hand side of Eq.~(\ref{eq:rho}), can be used to
calculate the Casimir energy as a function of the positions and
strengths $\alpha$ of the scatterers:
\beq
\label{eq:tote1d}
\cE=-\frac{1}{2\pi}\Im\int_0^\Lambda
dE\sqrt{E}\frac{\partial}{\partial E}\ln\det\Gamma(E+i 0^+,\{a\}).
\eeq

The interaction part of this energy is obtained by subtracting
from Eq.~(\ref{eq:tote1d}) the same quantity calculated with all
the $L_{ij}=|a_i-a_j|\to\infty$. In this limit $\Gamma$ becomes
diagonal (considering that $\Im E>0$) and the energy
(\ref{eq:tote1d}) becomes a sum of self-energies of isolated
objects. The self-energy of an isolated brane contains all the
usual ultra-violet divergences of the Casimir energy and must be
treated with care \cite{Jaffe:2003ji}. The sharp $\delta$-function
limit $r\to 0$ and the strong potential limit (sometimes called
\emph{the Dirichlet limit}) $V_0\to\infty$ are problematic for the
Casimir energy already in $d=1$ and the results depend on the
order in which the $r\to 0$, $V_0\to\infty$ and $\Lambda\to\infty$
limits are performed.\footnote{In this paper we are adopting a
cutoff regularization of the quantum field theory. Different
regularization schemes (like the widely used zeta-function, for
example) give different results for the divergences. The finite
part of the energy is, however, the same.} To avoid this problem
we will consider the situation in which the potential $V(x)$ is
finite, sufficiently smooth and localized over a finite distance
$0<r\ll L$, where $L$ is the separation between the branes.
Physically, this means that we assumed the cutoff $\Lambda$ of the
field theory to be much larger than any other momentum (square)
scale, like $V_0$ or $1/r^2$. The divergences that then arise in
the Casimir energy are entirely local: they will not affect the
interaction energy. The procedure of subtracting the self-energies
as described above then leaves a well-defined, finite interaction
energy between the branes. After this subtraction is performed one
can take the appropriate limits for the potential $r\to 0$ and
$V_0\to\infty$ and, provided the interaction energy remains finite
as we prove below, the resulting interaction energy is unique.

After performing the subtraction of the self-energies, the
interaction energy can be written as
\beq
\label{eq:manydelta1d}
\cE=-\frac{1}{2\pi}\Im\int_0^\Lambda
dE\sqrt{E}\frac{\partial}{\partial E}\ln\frac{\det\Gamma(E+i
0^+,\{a\})}{\det\Gamma(E+i 0^+,\infty)}.
\eeq
We keep using $\cE$ to indicate the interaction energy, confident
that this will not generate any confusion, since we will no longer
be interested in the total energy. The integrand in
Eq.~(\ref{eq:manydelta1d}) falls exponentially fast on the
semicircle $|E|\to\infty$ of the complex $E$ plane\footnote{In
particular it goes to zero like
$e^{-2\sin(\frac{\theta}{2})\sqrt{|E|}L}$ on the ray
$E=|E|e^{i\theta}$, $\pi>\theta>0$, where $L=\min|a_i-a_j|$ for
any $d$.} which allows us to integrate by parts, Wick-rotate to
the negative $E$ axis\footnote{During the Wick rotation we do not
pick any pole contribution on the positive imaginary semi-plane of
the first Riemann sheet because the total Hamiltonian
Eq.~(\ref{eq:H}) is self-adjoint.} and send the cutoff
$\Lambda\to\infty$. We can moreover remove the $\Im$ because all
the quantities are real and positive on the negative real $E$ axis
(since the propagator $\cG_0$ is real and positive for $E$ real
and below the spectrum) except for $\sqrt{-E+i0^+}=i\sqrt{E}$.

This leads us to a final, compact expression for $\cE$
\beq
\label{eq:Eintwickrot}
\cE=\frac{1}{4\pi}\int_0^\infty
\frac{dE}{\sqrt{E}}\ln\frac{\det\Gamma(-E,\{a\})}{\det\Gamma(-E,\infty)}.
\eeq
This is the main result of this paper and together with the
definition of $\Gamma$, Eq.~(\ref{eq:gammad3}), can be used to
calculate the interaction energy of point-like scatterers due to
fluctuations of the field $\phi$. In the rest of this paper we
present several examples of the applications of this formula.

\section{Examples}

As a first example and a check for our result,
Eq.~(\ref{eq:Eintwickrot}), let us calculate the well-known
interaction energy between two delta functions at distance $L$, in
1 dimension (we assume $\alpha_1=\alpha_2\equiv\alpha<0$):
\beq
\label{eq:EN1delta1d}
\cE=\frac{1}{4\pi}\int_0^\infty
\frac{dE}{\sqrt{E}}\ln\left(1-\frac{e^{-2L\sqrt{E+m^2}}}{(1-2\alpha\sqrt{E+m^2})^2}\right).
\eeq
This formula reproduces the usual results for the Casimir energy
of two penetrable plates in 1 dimension \cite{Mostepanenko}.

\begin{figure}
\centering
\hspace{-2cm}\includegraphics[width=10cm]{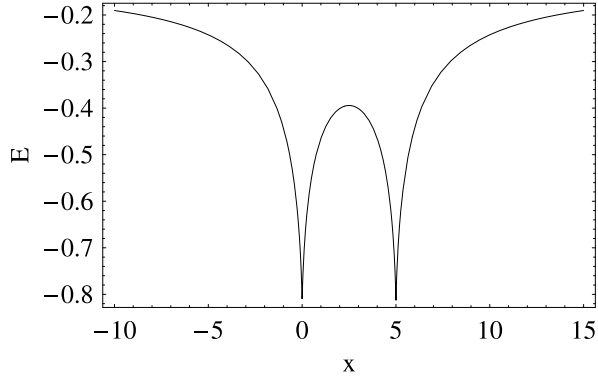}
\caption{\label{fig:3deltas} The interaction energy, in arbitrary
units, for three delta functions on the line as a function of the
position $x$ of one of them and $m=0$. $\alpha=-1$ for all three
deltas, one delta is held fixed at $x=0$, another at $x=5$.}
\end{figure}

As another example in $d=1$ consider the case of three repulsive
delta functions ($\alpha_i=-1$ for $i=1,2,3$). The interaction
energy can be calculated with the ease with which one can take a
determinant of a 3 by 3 matrix. The result is plotted in
Fig.~\ref{fig:3deltas} as a function of the position $x$ of one of
the three deltas while the other two are held fixed at $x=0$ and
$x=5$. The interaction energy is not additive: the interaction
energy of $N$ semi-penetrable plates does not split into a sum of
$N(N-1)/2$ terms due to pairwise interactions. Rather, by
expanding the logarithm a \emph{reflection expansion} is obtained
in the spirit of Ref.~\cite{optical}.

Before calculating the interaction energy for 2 or more deltas in
$d>1$ it is instructive to look at the case of a single delta
function centered in $x=0$, to introduce some properties of the
bound state of a single delta. Consider the case $d=3$. The
propagator is (for $\Im\ E>0$)
\beq
\label{eq:bound3d}
\cG(x',x;E)=\frac{e^{i|x'-x|\sqrt{E-m^2}}}{4\pi|x'-x|}+\frac{1}{\alpha-\frac{i\sqrt{E-m^2}}{4\pi}}\frac{e^{i(|x'|+|x|)
\sqrt{E-m^2}}}{16\pi^2|x'||x|}.
\eeq
There is evidently a pole at $E=E_0$ such that
$\sqrt{E_0-m^2}=-i4\pi\alpha$. For $\alpha<0$ this is a real bound
state at $E_0=m^2-16\pi^2\alpha^2$ and the wave function $\psi_0$
of this bound state is obtained by noticing that
\beq
\cG\sim\frac{1}{E_0-E}\psi^*_0(x')\psi_0(x),
\eeq
for $E$ near the pole $E_0$. We hence expand (\ref{eq:bound3d})
about $E_0$ to find
\beq
\psi_0(x)=\frac{\sqrt{2(-\alpha)}}{|x|}e^{-4\pi(-\alpha)|x|}.
\eeq
For $\alpha>0$, on the contrary, the pole is on the $2^{nd}$
Riemann sheet and hence is a virtual state and does not belong to
the spectrum of $H$. Whether this pole is real or virtual,
physically it represents the $s$-wave scattering over a
concentrated attractive potential. $1/\alpha$ is indeed
proportional to the scattering length in the $s$-wave channel
\cite{Zeldovich}. The $s$-wave is the only contribution surviving
in the limit when the scatterer is small compared to the
wavelength $1/\sqrt{E-m^2}$.

We have to require the spectrum of $H$ to be contained in the
positive real axis for the vacuum, $\phi=0$, of our field theory
to be stable. So if $\alpha<0$ we have to choose
$m>4\pi(-\alpha)$. If $\alpha>0$ any choice of $m$, in particular
$m=0$, is enough to ensure the stability of the $\phi=0$
vacuum.\footnote{It is worth noting that the opposite choice for
the sign of $\alpha$ is needed to avoid a bound state in the $d=1$
case.}

\begin{figure}
\hspace{-2.5cm}\includegraphics[width=10cm]{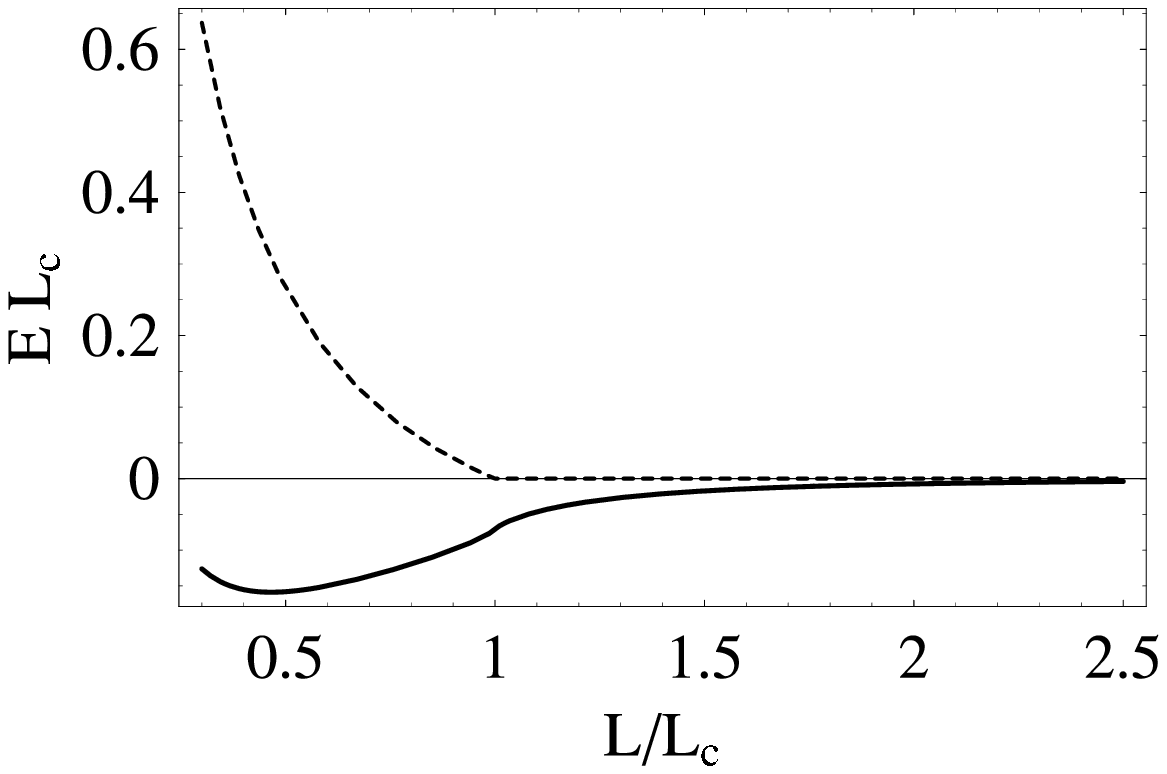}
\hspace{-2cm}\includegraphics[width=10.25cm]{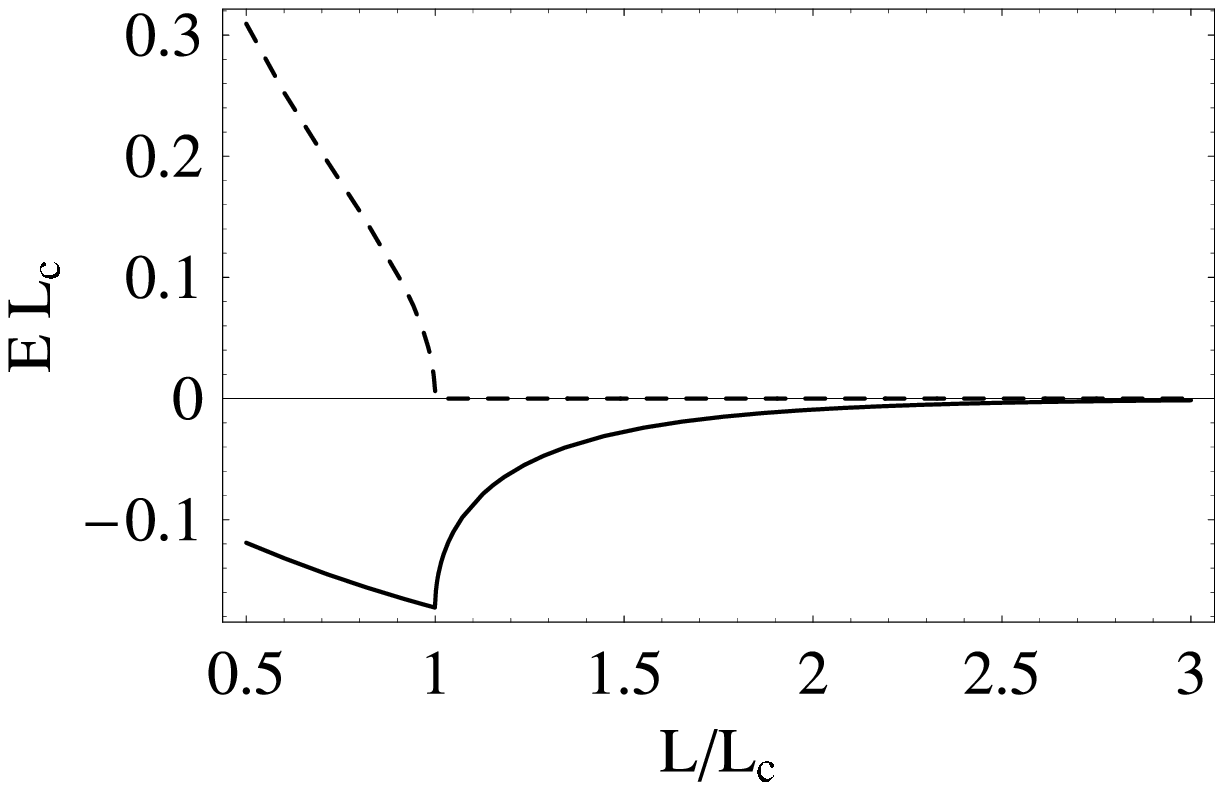}
\caption{\label{fig:2delta3d} Interaction energy $\cE$ (the
continuous line is $\Re\cE$ and the dashed line is $-\Im\cE$) in
units of $1/L_c$, for two delta functions as a function of their
distance $L$. a) The $\mathbb{R}^3$ case  with $m=0$.  b) The
$\mathbb{R}^2$ case with $m/M=2$.}
\end{figure}

Considering the case $d=3$ further, let us now calculate the
interaction energy between two identical delta functions with
$\alpha_1=\alpha_2=\alpha>0$ (so, according to the previous
paragraph no bound state exists for isolated scatterers) at a
distance $|a_1-a_2|=L$. After the Wick rotation and defining
$k\equiv\sqrt{E}$ we obtain
\beq
\label{eq:2deltasE3d}
\cE=\frac{1}{2\pi}\int_0^\infty
dk\ln\left(1-\frac{e^{-2L\sqrt{k^2+m^2}}}{
L^2\left(4\pi\alpha+\sqrt{k^2+m^2}\right)^2} \right).
\eeq
It is not difficult to see that it exists a critical distance
$L_c$, being the positive solution of the equation
\beq
L_ce^{mL_c}=\frac{1}{4\pi \alpha +m},
\eeq
such that if $L<L_c$ the argument of the logarithm in
Eq.~(\ref{eq:2deltasE3d}) becomes negative for sufficiently small
$k$ and we get a negative imaginary part in the Casimir energy. A
negative imaginary part of $\cE$ means, as usual, an instability
of the $\phi=0$ vacuum in the presence of the two $\delta$'s. In
fact, by studying the eigenvalues of the matrix $\Gamma$ one can
see that for $L<L_c$ the spectrum of $H$ has a bound state with
negative $E$, and since $E=\omega^2$ this is a clear indication
for the existence of a tachyon. We will return on the implications
of this instability for the low-energy physics.

The force $\cF\equiv-\partial\cE/\partial L$, always attractive
and central, diverges logarithmically at the critical length
$L_c$. For $m=0$ and $(L-L_c)/L_c\ll 1$ one finds:
\beq
\label{eq:divergeF3d}
\cF\simeq-\frac{1}{4\pi L^2}\ln\left(\frac{L_c}{L-L_c}\right).
\eeq

The long-distance behavior, $L\gg L_c$, of the force depends on
the mass of the boson $\phi$. For $m>0$ the potential between the
two $\delta$'s decreases exponentially. For $m=0$, instead, a
power-law tail is obtained:
\beq
\cE\simeq-\frac{L_c^4}{4\pi L^5}.
\eeq
This $1/L^5$ law is stronger than the Casimir-Polder law (induced
polarization interaction \cite{CasimirPolder}) which falls like
$1/L^7$. This means that we should not think of these delta
functions as mimicking polarizable molecules or metallic
particles. Indeed, to correctly describe a metallic sphere of
radius $R$, surface $\Sigma$ and penetration depth $r_0$ one
should rather assume that $r_0\ll R$ adding hence to $H_0$ in
Eq.~(\ref{eq:H0}) a potential $V(x)=\int_\Sigma d^{2} y
\mu\delta^{(3)}(x-y)=\mu\delta(r-R)$ and send $\mu\to\infty$
\emph{before} sending $R\to 0$. This clearly is a different limit
than the one we are describing here.

Now that we have discussed the divergences associated with
$\Gamma$, its renormalization and we know about the existence of
vacuum instabilities related to negative $E$ bound states of the
hamiltonian $H$ we are ready to tackle the two dimensional case
where all these complications arise at the same time.

The free propagator is $ \cG_0(x',x;E)=\frac{i}{4}
H_0^{(1)}(\sqrt{E-m^2}|x'-x|)$ if $\Im E>0$. Notice that even for
a single delta no choice of $\alpha$ eliminates the bound state.
There will always be at least one bound state with energy
$E_0=m^2-M^2 e^{-4\pi\alpha}$. This is due to the fact that any
attractive potential in 2 dimensions has a bound state. We must
choose our mass such that $E_0>0$ and the instability is not
present (it suffices that $m>Me^{-2\pi\alpha}$). However we will
see that in $d=2$, exactly as in the $d=3$ case discussed above,
in the presence of two or more $\delta$'s there exists a critical
distance such that for closer approach a bound state has $E<0$,
generating a tachyon again.

Take $N$ point-like scatterers, each with renormalized strength
$\alpha_i$ and a renormalization mass $M$. It is convenient to
define $M_i\equiv M e^{-2\pi\alpha_i}$ so $\Gamma$ in
(\ref{eq:GNdelta}) is
\beq
\Gamma_{ij}=\left(\frac{1}{2\pi}\ln
\frac{\sqrt{E-m^2}}{iM_i}\right)\delta_{ij}-\widetilde{\cG}_0(a_i,a_j;E).
\eeq

The interaction energy for two identical deltas ($M_1=M_2=\cM$,
and $m>\cM$ as required for the stability of isolated scatterers)
separated by a distance $L$ is ($k\equiv \sqrt{E}$)
\beq
\cE=\frac{1}{2\pi}\int_0^\infty
dk\ln\left(1-\frac{K_0^2(L\sqrt{k^2+m^2})}{\ln^2\left(\sqrt{k^2+m^2}/\cM\right)}\right),
\eeq
where $K_0$ is a Bessel function $K$ of order 0, and the critical
length is the solution of the equation $K_0(mL_c)=\ln m/\cM$. The
force diverges as $L\to L_c$ in $d=2$ as well but the explicit
expression is more difficult to recover. For $L\gg L_c$ the force
is exponentially small, since we had to assume a mass $m>0$ for
the field $\phi$.

\section{Localized Vacuum Instability}
\label{sec:Locinst}

Let us calculate the shape of the tachyon in $3$ dimension found
in the discussion after Eq.\ (\ref{eq:2deltasE3d})~(take $m=0$).
Let us first notice that (for any number of scatterers) the
Wick-rotated matrix $\Gamma(-E)$ is real and symmetric and can
hence be put in diagonal form. In the case at hand we have only
two delta functions with equal strength $\alpha$ and one can show
that the spectral decomposition of $\Gamma^{-1}$ is
\beq
(\Gamma^{-1})_{ij}(-E,\{a_1,a_2\})=\frac{1}{\gamma_1}
v^{(1)}_{i}v^{(1)}_{j}+\frac{1}{\gamma}_2 v^{(2)}_{i}v^{(2)}_{j}
\eeq
where
\barr
\gamma_1&=&-\frac{e^{-L{\sqrt{E}}}}
   {4\pi L}
   + \frac{{\sqrt{E}}}{4\,\pi } +
  \alpha,\\
\gamma_2&=&\frac{e^{-L{\sqrt{E}}}}
   {4\pi L}
   + \frac{{\sqrt{E}}}{4\,\pi } +
  \alpha,
\earr
and
\barr
v^{(1)}&=&\{\frac{1}{\sqrt{2}},\frac{1}{\sqrt{2}}\},\\
v^{(2)}&=&\{\frac{1}{\sqrt{2}},-\frac{1}{\sqrt{2}}\}.
\earr

The bound state pole is generated by a zero $E^*$ in the
$\gamma_1$ eigenvalue. For $L<L_c=1/4\pi\alpha$ we have $E^*$ as
the real positive solution of the equation $\gamma_1(E)=0$
(remember: the integration variable $E$ appears as $-E$ in
$\Gamma$ so positive $E$ here are real, negative eigenvalues of
$H$)
\beq
\sqrt{E}+\frac{1}{L_c}=\frac{e^{-L\sqrt{E}}}{L}.
\eeq
Comparing the behavior of the propagator for $E$ close to a pole
$E^*$
\beq
\cG(x',x;E)\simeq\frac{\psi^*_0(x')\psi_0(x)}{E^*-E}
\eeq
with
\beq
\cG(x',x;E)\simeq \frac{1}{\gamma_1}
\sum_{i,j}v_i^{(1)}v_j^{(1)}\cG_0(x',a_i;E)\cG_0(a_j,x;E)
\eeq
we find the wave function of the (not normalized) bound state as
\beq
\psi_0(x)=\frac{e^{-\sqrt{E^*}|x-a_1|}}{|x-a_1|}
+\frac{e^{-\sqrt{E^*}|x-a_2|}}{|x-a_2|}.
\eeq

Something can be said also in the case in which we have many
identical defects (and assume all the $a_{ij}$'s are of the same
order of magnitude), without necessarily having to solve the
equations explicitly. Instructed by the previous analysis we can
state that the ground state will be a highly symmetric state
$v\sim\{1/\sqrt{N},...,1/\sqrt{N}\}$ which will then give a
symmetric wave function $\psi_0(x)\propto\sum_i
v_{i}\cG(x,a_i;E^*)$ delocalized over the entire array of defects.
The positivity of the $v_i$'s coincide with the constraint that
the ground state must not have any node.

Hence for $L<L_c$ a free field theory coupled to these defects
does not make any sense. Its vacuum state $\phi=0$ is unstable.
The imaginary part of the energy (as an analytic continuation to
$L<L_c$) is related to the `decay time' of the vacuum state, due
to particle creation.

Let us for a moment speculate on the consequences of this
instability. Adding higher order terms in $\phi$ to the Lagrangian
---one can for example think of adding a $\lambda\phi^4$ term---
should eventually stabilize the field with a vacuum expectation
value (vev) $\overline{\phi}(x)\neq 0$ in a somewhat large region
around the two scatterers. However the actual value of the vev
$\overline{\phi}(x)$ and the size and shape of the condensation
region cannot be easily constructed and will be subject of future
work.

This scenario of a local condensation and creation of localized
vacuum instabilities due to defects could be interesting in
inflation cosmology as well (the field $\phi$ being the inflaton).
It must be also remarked that a similar scenario occurs in brane
cosmology when an open string has its ends attached to two
D-branes \cite{Quevedo}. When the branes are pushed closer than a
critical length one of the modes of the string becomes a tachyon.

\section{Extension to $n-1$ transverse dimensions}
\label{sec:Extn}

Now that we know the density of states $\rho(E)$ for the `basic'
problem of points in 1, 2 and 3 dimensions, we can move along the
lines of Table I to generate solutions for manifolds with
co-dimensions 1, 2 and 3. We shall then add $n-1$ transverse, flat
dimensions. The total dimension of the space is now $d+n-1$. The
calculations in the preceding part of this paper can be recovered
by putting $n=1$ in all the formulas. We will use the methods of
\cite{JaffePRL} and \cite{Buoyancy} where one solves for the
density of states $\rho(E)$ of the basic problem on a
$d-$dimensional section and insert the result in the equation for
the energy per unit $n-1$-dimensional `area' $S$ \cite{JaffePRL}
\beq
\cE^{(n)}=\int_{\mathbb{R}^{n-1}}\frac{d^{n-1}p}{(2\pi)^{n-1}}\int_0^\infty
dE \frac{1}{2}\left(\sqrt{p^2+E}-\sqrt{p^2}\right)\rho(E).
\eeq
The subtraction $-\sqrt{p^2}$ removes a divergent but
$a-$independent term, since the integral $\int dE\rho(E)$ is
$a$-independent. We will also remove the $a$-independent
`self-energy' terms by subtracting from $\rho(E)$ the density
$\rho(E,\infty)$ with all $a_{ij}\to\infty$. We can then perform
the (dimensionally regularized) integral over $p$, Wick-rotate and
perform an integration by parts on $E$ to obtain
\begin{equation}
\cE^{(n)}=\frac{1}{2\pi}\frac{\Gamma(1-\frac{n}{2})\sin{\frac{n\pi}{2}}}
{(4\pi)^{n/2}}\int_{0}^{\infty}dE
E^{\frac{n}{2}-1}\ln\frac{\det\Gamma(-E;\{a\})}{\det\Gamma(-E;\infty)}.
\label{eq:transverse1}
\end{equation}
For example the interaction energy (per unit length) of two
straight, infinite strings in $3$ dimensions ($d=2, n=2$ so
$d+n-1=3$) put at a distance $L$ is
\beq
\cE^{(2)}=\frac{1}{8\pi}\int_0^\infty
dE\ln\left(1-\frac{K_0^2(L\sqrt{E+m^2})}{\ln^2\left(\sqrt{E+m^2}/M\right)}\right).
\eeq
One can hence calculate the interaction energy of any two flat
manifolds due to the quantum fluctuations of a bulk field $\phi$.
As an example in co-dimension 1 consider the Randall-Sundrum
scenario \cite{RS} with two branes at a distance $r_c$ from each
other. The fluctuations of a given component of the metric $G$ or
of a bulk field $\phi$ (see \cite{Hofmann:2000cj} and references
therein), have a space-dependent mass with two delta functions
singularities on the two branes. The attractive force due to the
quantum fluctuations of this field has a Casimir-like behavior.
The curvature in the $5^{th}$ direction does not change the
physics.\footnote{If the brane is inside an horizon for the 5d
metric most probably this assertion is not true, however. But this
is not the case for the Randall-Sundrum model.} If however the
branes have co-dimension 2 or 3 (and are defined as the limit of
an attractive potential) is in the class of problems that we have
studied in this paper and a perturbation would eventually
condense, if $r_c<L_c$. Following the same arguments above we can
also say that if the branes have co-dimension $>4$ the
fluctuations in the bulk will not see the brane. The cosmological
implications of such a scenario will be subject of future work.

\section{Omissions and Applications}
\label{sec:omiss}

The propagator, Eq.~(\ref{eq:GNdelta}), comes directly from
scattering theory. In that context it was natural to assume that
the interaction between the particle and the scatterer (consider
Fermi \cite{Fermi} and Zel'dovich \cite{Zeldovich} examples) is
attractive. One considers an attractive center whose attraction
grows when $r_0\to 0$ such that at most one bound state is present
and its energy remains finite (\emph{i.e.}\ of $\Ord{1}$). Even
thought we assumed that only one bound state is present at
energies of $\Ord{1}$ this is the most generic situation that can
occur in scattering theory. In fact if a second bound state is
present, it will be an energy $\Ord{1/r_0^2}$ below our bound
state. In the limit $r_0\to 0$ its influence on low-energy
scattering disappears. It goes out of the spectrum. In scattering
theory however such a negative energy state is harmless. This is
not the case for a bosonic field, for which it represents a
tachyon.

One may wonder: what happens if the potential is \emph{repulsive}
and concentrated? The answer is that for $d>1$ its influence on
the scattering matrix (and hence on the spectrum) disappears when
$r_0\to 0$. Obviously this is not true in 1 dimension because we
cannot `go around' the scatterer. For a repulsive potential in
$d>1$ the renormalization procedure leading to (\ref{eq:GNdelta})
cannot be performed since $\mu$ has the wrong sign and sending
$r_0\to 0$ just kills the correction to $\cG_0$ in
(\ref{eq:GNdelta}).

More precisely, if in the Lagrangian we include a term
$V_0\theta(r_0-|x|)\phi^2(x)$ with $V_0>0$ and then we take the
limit $V_0\to\infty$ and $r_0\to 0$ with $V_0r_0^2$ finite, the
spectrum we obtain is just the free one: the scatterer disappears.
In a sense, the only smile the Cheshire cat can leave behind is
the lightly (\emph{i.e.}\ $\Ord{1}$ instead of $\Ord{1/r_0^2}$)
bound (or virtual) state. If this is not present then the
scatterer is invisible to the fluctuations.\footnote{We have
already remarked on the impossibility to consider the
Casimir-Polder interaction between small metallic spheres as the
interaction of point-like scatterers the way we construct them
here. Small metallic particles of radius $r$ still have a
penetration depth $r_0\ll r$ so effectively are co-dimension 1
surfaces.} If the purpose of calculating the effective action was
to calculate quantum corrections to a classical solution (as often
occurs) then we deduce that for a repulsive potential or for
$d\geq 4$ the classical solutions are unchanged by quantum
fluctuations.

Let us now comment on two possible applications of the formalism
we have developed: cosmic strings and concentrated Aharonov-Bohm
fluxes. We anticipate that further work is required in both cases.
In the literature on cosmic strings the difficulty generated by a
bound state tied to a \emph{single} cosmic string has been
recognized a long time ago \cite{Allen,Kay}. In that context the
bound state arising from Eq.~(\ref{eq:1delta1dren}) is rightly
considered fictitious, because the smoothed potential is always
positive ($\mu>0$). Nonetheless, in \cite{Allen} after projecting
out this bound state at $E_0<0$, the propagator (\ref{eq:GNdelta})
is trusted and shown to be in good agreement with the numerical
solution of the smoothed problem. It is not clear if projecting
out a state from the propagator by hand has non-trivial (wrong)
consequences on the density of states and the Casimir energy so we
preferred not to follow this path even if it gives correct results
for other quantities. We hence required the field to have a
non-zero mass so that this bound state is stable. In the end it is
not clear if the Casimir attraction and the birth of the tachyon
could arise in cosmic strings coupled with bulk fields.

Another example to which the above techniques and results should
be relevant is the case of a fermion around a concentrated tube of
flux (Aharonov-Bohm case\footnote{The one loop energy of QED flux
tubes has been calculated in \cite{Graham:2004jb} using a
combination of analytical and numerical methods. Our method could
be used to calculate the interaction energy of two such tubes in
the limit where their radius is small compared to their relative
distance.}). The spectrum of Dirac's equation can be inferred from
that of a Klein-Gordon equation after squaring the former. The
fact that we are dealing with fermions rather than bosons, is not
a difficulty. Another difficulty however arises: for Aharonov-Bohm
fluxes and the more general case of cosmic strings charged under
some U(1) symmetry, it has been shown \cite{AW} that the
contribution to the scattering cross section given by the non-zero
external vector potential is asymptotically larger in the low
energy regime than the contribution of the singularity in the
core. Since we believe that cross sections and Casimir forces are
tightly bounded quantities we would not apply any of the above
arguments without treating the propagation in the external space
properly. This will be done elsewhere.

Renormalization of branes coupling for a single brane (or
$\delta$-function in our case) with co-dimension 2 (and dimension
5) in a conical space has been studied in \cite{GW1}.\footnote{See
also \cite{Naylor:2002xk}.} Arising from local divergences, the
renormalization flow is not affected by the presence of other
branes and the results in \cite{GW1} apply also to our situation.
Their renormalization of the brane coupling $\mu$ ($\lambda_2$ in
their notation) corresponds to our renormalization
$\alpha\to\alpha_r$. Their renormalization of the effective action
corresponds to our subtraction of the $a$-independent terms in the
Casimir energy. These two are the only subtractions needed (if
$\phi^4$ terms are not present) and it is heartening to see that
our results coincide with those of \cite{GW1}. Moreover one can
make an amusing observation if one compares the two approaches to
the delta function, the one in terms of scattering (that we used
here) and the one in \cite{GW1} in terms of renormalization group.
Notice that the renormalization group flow for $\mu$ is IR free
and has a Landau pole: the location of the Landau pole coincides
with the location of the bound state in our approach.

\section{Conclusions}

We have calculated the force between an arbitrary number of
surfaces (branes) with co-dimension $>1$ due to the quadratic
fluctuations of a boson $\phi$ living in the bulk. The force turns
out to be attractive and it diverges when the distance between the
branes approaches a critical value $L_c$. This phenomenon has no
analogues in the widely studied co-dimension 1 case.

The divergence of the force is accompanied by the birth of a
vacuum instability, a mode with negative mass squared localized
around the scatterer. In 3 dimensions, the long-range properties
of this force (decreasing like $1/L^6$) are shown to be different
from the Casimir-Polder $1/L^8$ law, the explanation relying in
the proper mathematical definition of the point-like limit.

Some implications of these effects have been pointed out.

\section{Acknowledgments}

I would like to thank E.~Fahri, J.~Goldstone, A.~Guth, A.~Hanany,
R.~Jackiw, H.~Liu and C.~Nunez for discussions and in particular
R.~Jaffe for many discussions and for a critical reading of the
manuscript. This work has been supported in part by funds provided
by the U.S.~Department of Energy (D.O.E.) under cooperative
research agreement DE-FC02-94ER40818. I acknowledge partial
financial support from INFN through a Bruno Rossi fellowship.

\end{document}